\documentclass[fleqn,10pt]{wlscirep}
\usepackage[utf8]{inputenc}
\usepackage[T1]{fontenc}

\title{Joint Optimization of Electric Vehicle Routes and Charging Locations through Learning Charge Constraints Using QUBO Solvers}

\author[1,*]{Akihisa Okada}
\author[1]{Keisuke Otaki}
\author[1]{Hiroaki Yoshida}
\affil[1]{Toyota Central R\&D Labs., Inc., Tokyo, 112-0004, Japan}

\affil[*]{a-okada@mosk.tytlabs.co.jp}

\begin{abstract}
Optimal routing problems of electric vehicles (EVs) have attracted much attention in recent years, and installation of charging stations is an important issue for EVs. Hence, we focus on the joint optimization of the location of charging stations and the routing of EVs. When routing problems are formulated in the form of quadratic unconstrained binary optimization (QUBO), specialized solvers such as quantum annealers are expected to provide optimal solutions with high speed and accuracy. However, battery capacity constraints make it hard to formulate into QUBO form without a large number of auxiliary qubits.
Here, we propose a sequential optimization method utilizing the Bayesian inference and QUBO solvers, in which the battery capacity constraints are automatically learned. This method enables us to optimize the number and location of charging stations and the routing of EVs with a small number of searches. 
Applying this method to a routing problem of 20 locations, 
we observed consistent convergence toward battery-feasible solutions across independent runs, demonstrating stable learning behavior of the proposed framework.
Small-scale validation experiments using exhaustive enumeration show that the framework reliably discovers feasible configurations close to the global optimum, while runtime and QUBO-size analyses clarify its computational characteristics.
\end{abstract}
\begin{document}

\flushbottom
\maketitle

\thispagestyle{empty}

\section*{Introduction}
To promote a sustainable society, it is essential to improve the efficiency and practicality of electric vehicles (EVs). 
Although the adoption of EVs is steadily progressing, their limited driving range and the insufficient availability of charging infrastructure remain significant barriers to widespread use~\cite{Rauh2015, Coffman2017, Pamidimukkala2023}. This issue becomes even more pressing in logistics and delivery services, where vehicles must travel to multiple locations. 
Therefore, it is necessary to consider route optimization that takes battery levels into account, as well as determine the optimal number and placement of charging stations.

Route planning for EVs can be effectively modeled as a type of combinatorial optimization problem.
In particular, it is formulated as a capacitated vehicle routing problem (CVRP)~\cite{Schneider2014, Afroditi2014, keskin2016, Montoya2017}, which adds cargo capacity constraints to path finding problems called vehicle routing problems.
This is because battery capacity can be treated similarly to cargo capacity: both must be managed to stay within a certain range throughout the route.
Based on this analogy, various studies have been conducted on route optimization for EVs. There are studies that optimize the placement and selection of charging stations~\cite{Lam2014, Moghaddam2018}.

In recent years, the quadratic unconstrained binary optimization (QUBO) formulation has attracted much attention as an effective method for solving combinatorial optimization problems. 
QUBO problems can be efficiently and accurately solved using quantum annealers~\cite{Johnson2011} and other specialized solvers~\cite{Yamaoka2015, Tatsumura2019, Matsubara2020}. 
This approach has been successfully applied in various domains, including polymer phase separation~\cite{Endo2022}, mobility-on-demand services~\cite{Otaki2023}, 
{financial optimization~\cite{Ding2023}, trajectory deconfliction~\cite{Stollenwerk2020}, and facility-location optimization problems~\cite{Ciacco2026}.} 
Broader applications across many scientific and industrial fields have also been reported~\cite{Yarkoni2022, Hussain2020, Inoue2021, Nishimura2019, Ohzeki2019, Syrichas2017, Tabi2021, Weinberg2023, Ye2023}.
{
These studies demonstrate that QUBO-based approaches can be applied to a wide range of NP-hard optimization problems.  
Comprehensive surveys of quantum and quantum-inspired optimization approaches further discuss the broader landscape, including their potential and limitations for constrained optimization problems~\cite{Li2020,Gharehchopogh2022}.
} 
Therefore, by formulating the capacitated vehicle routing problem in the QUBO framework, high-precision and rapid solutions can be expected.

Although several studies have formulated the CVRP within the QUBO framework~\cite{Feld2019, Irie2019, Weinberg2023}, applying it to battery constraints remains challenging. 
For instance, Feld et al.~\cite{Feld2019} proposed a QUBO-based route optimization approach for the CVRP, where vehicle capacity is represented as a discrete variable using one-hot encoding. 
However, when this capacity is reinterpreted as battery level (a continuous quantity that changes through movement and charging) binary approximation with high resolution becomes necessary.
Thus, a large number of bits is required, which leads to reduce the accuracy of solution or exceed the upper limit of bits that solver can handle.
More concretely, when battery level is discretized with resolution $\Delta$, 
the number of battery levels becomes $R \sim Q_\mathrm{max} / \Delta$, where $Q_\mathrm{max}$ denotes the maximum battery capacity.
If battery states are represented at each visit step $t = 1, \ldots, N$ using one-hot encoding, the number of additional binary variables scales as $O(NR)$.
For example, in practical EV routing scenarios, modeling battery levels with even modest precision (e.g., 1\% resolution) can easily yield several hundred discrete levels. 
When such states are introduced at every visit step, the number of battery-related binary variables can rapidly dominate the routing variables, significantly increasing QUBO size and coupling density.
Although binary encoding can reduce the number of state variables to $O(N\log R)$, enforcing sequential energy updates (arrival $\to$ charging $\to$ consumption), bound constraints, and nonlinear charging behavior typically introduces additional auxiliary variables and dense quadratic penalty terms. 
These scaling and constraint-density issues represent fundamental challenges in explicit QUBO formulations of electric vehicle routing problems (EVRP) and CVRP.
Consequently, improving the fidelity of battery modeling directly translates into larger and denser QUBO formulations.
Moreover, even formulating route optimization that considers only battery constraints is complex. 
Extending the model further to simultaneously optimize the placement of charging stations, whose locations significantly affect routing, poses an even more intricate and demanding challenge within the QUBO framework.

In this study, we propose a black-box optimization approach to effectively solve the coupled problem of charging station placement and battery-constrained routing. 
We model the relationship between charging station placement and the resulting optimal route as a black-box function, which cannot be easily formulated in advance. 
To tackle this, we adopt an iterative optimization strategy that explores the search space and gradually learns black-box function to improve the solutions. 
Recently, techniques have been developed to incorporate such black-box functions directly into the QUBO framework, enabling integration with QUBO solvers like quantum annealers during the optimization process~\cite{Kitai2020, Kadowaki2022, Inoue2022, Matsumori2022, Okada2023}. 
In our approach, we use BOCS (Bayesian Optimization of Combinatorial Structures)~\cite{Baptista2018}, a Bayesian inference-based method for learning scheme.
This approach allows us to avoid complex mathematical modeling of battery constraints by using a simplified penalty term for constraints.
By learning the mapping between station placements and energy-efficient routes, our method efficiently identifies high-quality solutions with only a limited number of evaluations.

While recent methods have made progress in formulating constrained combinatorial optimization problems in the QUBO framework, they still involve complexity. 
For example, Shirai et al.~\cite{Shirai2024} proposed an approach that optimizes the annealing schedule to obtain constraint-satisfying solutions. 
However, such methods require explicit constraint modeling and additional post-processing steps to recover valid solutions. 
In contrast, our proposed black-box learning method bypasses these complexities by eliminating the need for constraint formulation and allowing efficient exploration of feasible solutions through learning.

\section*{Methods}
\subsection*{Problem Setting}
\begin{figure}
\centering
\begin{minipage}{0.8\linewidth}
\centering
\includegraphics[width=8cm]{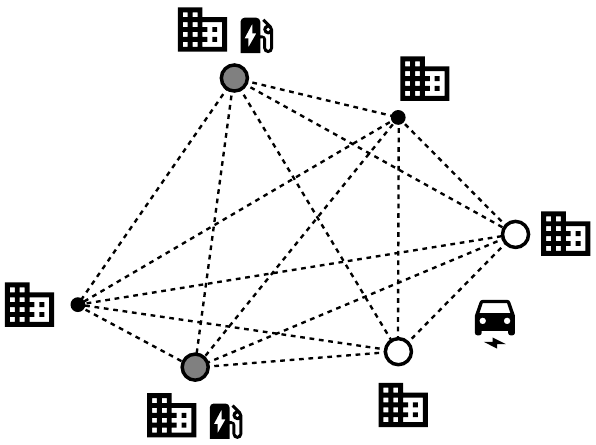}
\caption{Problem setting. Locations where charging stations cannot be installed are represented as black dots, while locations where installation is possible are shown as circles. The objective is to determine a vehicle route that visits all required locations while maintaining a positive battery level throughout the tour and minimizing total travel cost. Simultaneously, the placement of charging stations must be determined—indicated by gray circles for selected locations and white circles for unselected ones. Dashed lines represent candidate segments of the tour.}
\label{fig:setting}
\end{minipage}
\end{figure}

To evaluate the effectiveness of our proposed framework for jointly optimizing the EV route and charging station placement under energy constraints, we consider a scenario involving a single EV with a maximum battery capacity of $Q_\mathrm{max}$ tasked with visiting $N$ locations. 
Among these locations, up to $M$ may be selected for the installation of charging stations, each of which allows the EV to recharge by a fixed amount $Q_\mathrm{charge}$ (see Fig.~\ref{fig:setting}). The energy cost of traveling from location $i$ to $j$ is denoted by $C_{ij}$, which can be negative in some cases due to regenerative braking effects on downhill routes. Moreover, the energy cost is asymmetric, so $C_{ij} \neq C_{ji}$ in general. Given these conditions, our objective is to jointly determine the optimal placement of charging stations and the EV’s route such that the total energy consumption is minimized while satisfying the battery constraints throughout the journey.

\subsection*{Joint optimization of Charging Station Placement and Vehicle Routing Using Black-Box Optimization}
\begin{figure}[ht]
\centering
\begin{minipage}{0.8\linewidth}
\centering
\includegraphics[width=5cm]{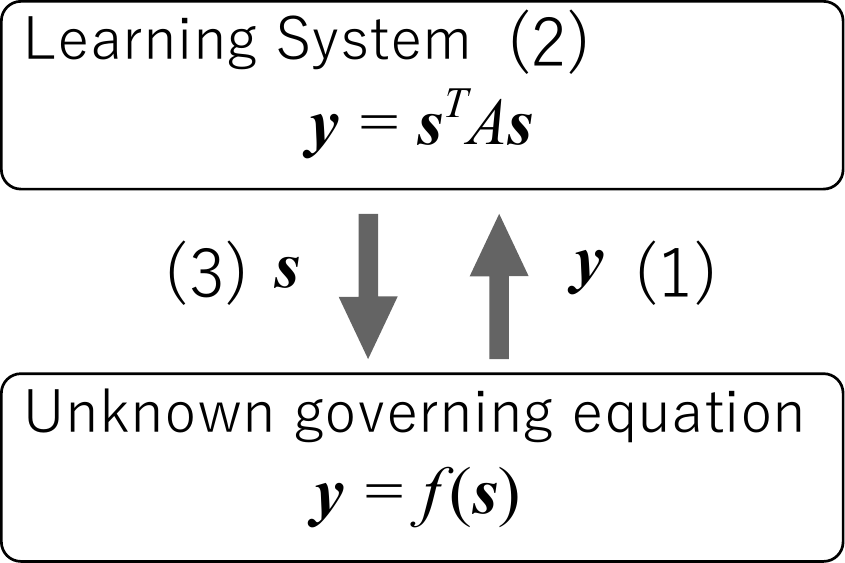}
\caption{Optimization Cycle Using BOCS.
The BOCS algorithm performs sequential optimization for a system in which the relationship between the binary input vector $s$ (a bit string) and the corresponding output $y$ is unknown. 
The process consists of the following three steps, which are repeated iteratively:
(1) Obtain a data pair ($s$, $y$) by evaluating the output $y$ for a given input $s$; 
(2) Learn the relationship between inputs 
$s$ and outputs $y$ in the form of a QUBO model using all previously collected data pairs;
(3) Based on the learned model, determine the next candidate input $s$ that is expected to yield an optimal output $y$.}
\label{fig:bocs}
\end{minipage}
\end{figure}

To lay the groundwork for our proposed framework, we first outline the general process of black-box optimization using the QUBO formulation, as illustrated in Fig.~\ref{fig:bocs}. 
This approach is useful when the relationship between an input $s$ and its corresponding output $y$ is unknown or difficult to model explicitly. 
The black-box optimization process proceeds in three main steps: (1) obtain input-output data pairs ($s$, $y$) through experiments or simulations, (2) approximate the relationship between $s$ and $y$ by learning a QUBO model in the form $y = s^\top A s$, where $A$ is a coefficient matrix, and (3) use the learned model to determine the next input $s$ that is expected to yield a better output $y$. 
By repeating this cycle and updating the matrix $A$ with each new data point, the system gradually improves its prediction of optimal or near-optimal inputs. 
In step (3), optimization techniques such as Simulated Annealing (SA) or Quantum Annealing (QA) can be used to solve the QUBO and select the next input. 
One full execution of steps (1) through (3) is treated as a single search iteration. 
The BOCS (Bayesian Optimization of Combinatorial Structures) method performs this cycle for a fixed number of iterations, denoted as $N_\mathrm{search}$, starting with an initial dataset of $N_\mathrm{Init}$ input-output pairs used to construct the initial model.
This iterative process forms the foundation for the framework we develop in this section.

To tackle the vehicle routing problem with battery constraints, we formulate it as an above-mentioned black-box optimization problem and propose a framework that maps the input $s$ to the placement of charging stations and the output $y$ to the total tour cost, as illustrated in Fig.~\ref{fig:bocs}. 
The objective is to identify the optimal pair $(s, y)$ that minimizes energy consumption while satisfying battery constraints. 
Our framework consists of two key components: (1) determining the optimal route for a given station placement using a simplified and soft battery constraint, and (2) assigning a penalty value $b$ to evaluate constraint violations. 
Together, these components provide the evaluation value $y$ for any given placement $s$, enabling the black-box optimization process.
The optimal solution $s$ obtained through this framework corresponds to the optimal placement of charging stations, and the resulting route represents the optimal path for the EV. The following subsections provide detailed explanations of components (1) and (2).

\subsection*{Optimal route for a given station placement using a simplified and soft battery constraint} \label{sec:stgiven}
To model the vehicle routing problem with battery constraints in the QUBO framework, it is essential to convert key aspects of the problem into binary variables. 
Specifically, we need to represent the vehicle's route, express changes in battery charge due to travel and recharging, and define the associated costs, all of which are expressed using binary encoding. 
In addition, a cost function must be constructed to encode the battery capacity constraints, ensuring that battery levels remain within feasible bounds throughout the route. 
In this subsection, we lay the foundation for this formulation by first introducing the necessary binary variables and defining the core cost function related to vehicle routing.

To determine the visiting order, we introduce a discrete time step $t$, where $t = \{ 1, \cdots, N\}$, given that there are $N$ locations to be visited.
Next, to formulate the routing problem in QUBO form, we define binary variables that represent the tour. 
Let $x_{ijt}$ be a binary variable that takes the value 1 if the vehicle moves from location 
$i$ to location $j$ at time step $t$, and 0 otherwise.
The total cost associated with the tour, expressed in terms of battery consumption, is given by:
\begin{equation} 
\sum_{i, j, t = 1}^{N} C_{ij} x_{ijt}. \label{eq:move} 
\end{equation}
However, it is necessary to enforce constraints such that each location is visited exactly once and in a valid sequence. 
To express the objective function as a QUBO, these constraints are incorporated as penalty terms.
The constraint that each location must be visited exactly once is expressed as:
\begin{equation} 
\sum_{i=1}^N \left( \sum_{j, t = 1}^{N} x_{ijt} - 1 \right)^2. \label{eq:locone}
\end{equation}
Additionally, to ensure that the tour proceeds logically from one location to the next, the arrival location at one time step must match the departure location at the next. 
This continuity constraint is formulated as:
\begin{equation} 
\sum_{t=1}^{N} \sum_{j=1}^N \left( \sum_{i=1}^{N} x_{ijt} - \sum_{k=1}^{N} x_{j k (t+1)} \right)^2. \label{eq:flowcons}
\end{equation}
We impose cyclic indexing such that time step $N$ is followed by time step $1$.
Unlike link-based formulations in which binary variables independently represent edges, the present formulation employs a time-indexed arc-flow encoding.
Each time step $t$ selects exactly one outgoing arc, and the flow conservation constraint in Eq. \eqref{eq:flowcons} enforces continuity between consecutive time steps.
Since the time index runs from $t = 1$ to $N$ and each location must appear exactly once due to Eq. \eqref{eq:locone}, the resulting structure 
forms a single Hamiltonian cycle under sufficiently large penalty coefficients.
Therefore, additional subset-based subtour elimination constraints are not required.
If we wish to fix the departure location at time step $t=1$ to a specific location {$i_0$}, we can include the following penalty term:
\begin{equation} 
\sum_{i \neq {i_0}} \left( \sum_{j=1}^{N} x_{ij1} \right)^2. 
\end{equation}
Therefore, the objective function $\mathcal{H}_\mathrm{TSP}$, which considers only the minimization of the tour cost, is given by the following expression.
\begin{equation}
\mathcal{H}_\mathrm{TSP} = \sum_{i, j, t = 1}^{N}  C_{ij} x_{i j t} + \lambda_1 \sum_{i=1}^N \left( \sum_{j, t = 1}^{N}  x_{i j t} - 1 \right)^2 + \lambda_2 \sum_{t=1}^{N-1} \sum_{j=1}^N \left( \sum_{i=1}^{N} x_{i j t} - \sum_{k=1}^{N} x_{j k (t+1)} \right)^2 + \lambda_3 \sum_{i \neq {i_0}} (\sum_{j=1}^{N}  x_{i j 1})^2.
\end{equation}
Here, $\lambda_1$, $\lambda_2$, and $\lambda_3$ are sufficiently large penalty coefficients introduced to enforce the corresponding constraints.

Next, we introduce a method to represent battery levels using the defined binary variables. 
Let $s_i$ be a binary variable that takes the value $1$ if a charging station is installed at location $i$, and $0$ otherwise. 
Recall that $x_{ijt}=1$ indicates that the vehicle moves from node $i$ to node $j$ at step $t$, and is located at node $j$ at time step $t$.
To construct a QUBO-compatible objective, we introduce an approximate cumulative battery level $\tilde Q_t$ defined as
\begin{equation}
\tilde Q_t = Q_1 - \sum_{t^\prime=1}^{t-1} \left( \sum_{i, j =1}^N C_{ij} x_{i j t^\prime} - Q_\mathrm{charge} \sum_{i, j =1}^N s_i x_{i j t^\prime} \right).
\end{equation}
Here, $Q_1$ represents the initial battery level. 
It should be noted that $s_i$, whether a charging station is installed at location $i$ or not, is determined by the configuration proposed by BOCS. 
Therefore, in the current route optimization step, the variables $s_i$ can be treated as given.
Here, charging is associated with the departure node $i$ of each move.
The quantity $\tilde Q_t$ is used only within the QUBO formulation as a soft guidance term.

Although the battery level is constrained to lie between 0 and its maximum capacity, this study does not adopt a strict formulation using techniques such as slack variables. Instead, we propose a simplified approach that encourages the battery level to stay close to a reference value $Q_\mathrm{standard}$. 
For example, this reference may be set to half of the maximum battery capacity. 
This condition is incorporated into the model as a penalty term.
\begin{equation}
\mathcal{H}_\mathrm{Battery} = \sum_{t=1}^{N} ( \tilde Q_t - Q_\mathrm{standard})^2. \label{eq:battery_cost}
\end{equation}
Since $\tilde Q_t$ is linear in the binary variables, $H_{\mathrm{Battery}}$ is a quadratic polynomial and thus remains in the QUBO form.
Expanding Eq.~\eqref{eq:battery_cost} explicitly yields constant, linear, and quadratic terms in the binary variables. Therefore, the battery-guidance term preserves the QUBO structure without introducing higher-order interactions.
Note that $\tilde Q_t$ is an approximate battery representation used solely inside the QUBO formulation and does not enforce strict capacity feasibility.

Based on the above, a route that visits all locations while specifying the departure point and incorporating battery levels in a simplified manner can be obtained by minimizing the following QUBO-form objective function $\mathcal{H}_\mathrm{total}$.
\begin{equation}
\mathcal{H}_\mathrm{total} = \mathcal{H}_\mathrm{TSP} + \lambda_4  \mathcal{H}_\mathrm{Battery}. \label{eq:total_guide}
\end{equation}
The coefficient $\lambda_4$ is used to encourage the battery level $\tilde Q_t$ to remain close to the reference value $Q_\mathrm{standard}$.

\subsection*{Penalty for Satisfying Battery Constraints}
The cost of the vehicle route obtained by minimizing the objective function $\mathcal{H}_\mathrm{total}$ in the previous section is not, in fact, the true cost we wish to evaluate when considering battery constraints. 
This is because $\mathcal{H}_\mathrm{total}$ includes a term that penalizes deviations from the reference battery level $Q_\mathrm{standard}$, which does not match the actual battery constraint. 
The true constraint is that the battery level must remain within the range [0, $Q_\mathrm{max}$] at all times.
Battery feasibility is evaluated outside the QUBO by a deterministic sequential update.
The update ordering in the evaluator is defined as consumption first (travel), followed by charging at the departure node of that move.
Let $Q_0 = Q_1$.
Given the realized solution $\bar{x}_{ijt}$, the battery state is updated step by step.
First, after travel consumption,
\begin{equation}
Q_t^{\mathrm{arr}} = Q_{t-1} - \sum_{i, j=1}^{N} C_{ij}\,\bar{x}_{ijt}.
\end{equation}
Then, if the departure node is equipped with a charging station,
\begin{equation}
Q_t = Q_t^{\mathrm{arr}} + Q_{\mathrm{charge}} \sum_{i, j=1}^{N} s_i\,\bar{x}_{ijt}.
\end{equation}
Note that associating charging with the departure node is a modeling convention used for consistency between the QUBO formulation and the evaluator; an equivalent formulation can be obtained by shifting the time index without affecting feasibility outcomes.
It is important to emphasize that travel-based consumption and charging increments are handled sequentially rather than simultaneously in the evaluator. If these effects were applied in a single aggregated update, unintended behaviors could occur. For example, the battery might temporarily drop below zero due to travel consumption but appear feasible after immediate charging, or conversely exceed $Q_{\max}$ due to charging and then return within bounds after consumption. The sequential update adopted here prevents such artificial cancellation effects and ensures consistent feasibility evaluation.
Constraint violations are checked at two stages:
\begin{itemize}
\item battery depletion or regenerative over-capacity after travel,
	i.e., $Q_t^{\mathrm{arr}}<0$ or $Q_t^{\mathrm{arr}}>Q_{\max}$,
\item over-capacity after charging, i.e., $Q_t>Q_{\max}$.
\end{itemize}

Therefore, the evaluation value $y$ for a tour under battery constraints is calculated using the following procedure:
\begin{enumerate} 
\item Minimize $\mathcal{H}_\mathrm{total}$ using a QUBO solver, such as Simulated Annealing (SA) or Quantum Annealing (QA), and obtain the binary variables $\bar{x}_{ijt}$, which determine the tour route. 
\item Using the obtained $\bar{x}_{ijt}$, compute the travel cost (i.e., total energy consumption) $a$ according to Equation~(\ref{eq:move}). 
\item For the constructed tour, calculate the battery level at each time step. 
\item Define a penalty function $y_\mathrm{penalty}(t)$, which takes a nonzero penalty value $y_\mathrm{penalty}$ if the battery constraint is violated at time step $t$, and zero otherwise. 
\item Sum the penalty values over all time steps: $b \equiv \sum_{t=2}^{N}y_\mathrm{penalty}(t)$. 
\item Compute the final evaluation value as $y=a+b$. 
\end{enumerate}

In the above evaluator, over-capacity states are penalized but the battery level is not saturated at $Q_{\max}$.
This relaxed treatment maintains consistency with the soft-guidance mechanism toward $Q_{\mathrm{standard}}$ in the routing-QUBO.
Although non-physical intermediate states may temporarily occur during the search,
the solutions reported as feasible are restricted to tours with zero penalty,
i.e., those satisfying $0 \le Q_t^{\mathrm{arr}} \le Q_{\max}$ and $0 \le Q_t \le Q_{\max}$ for all $t$.

\subsection*{Overall Framework}
\begin{figure}[ht]
\centering
\includegraphics[width=12cm]{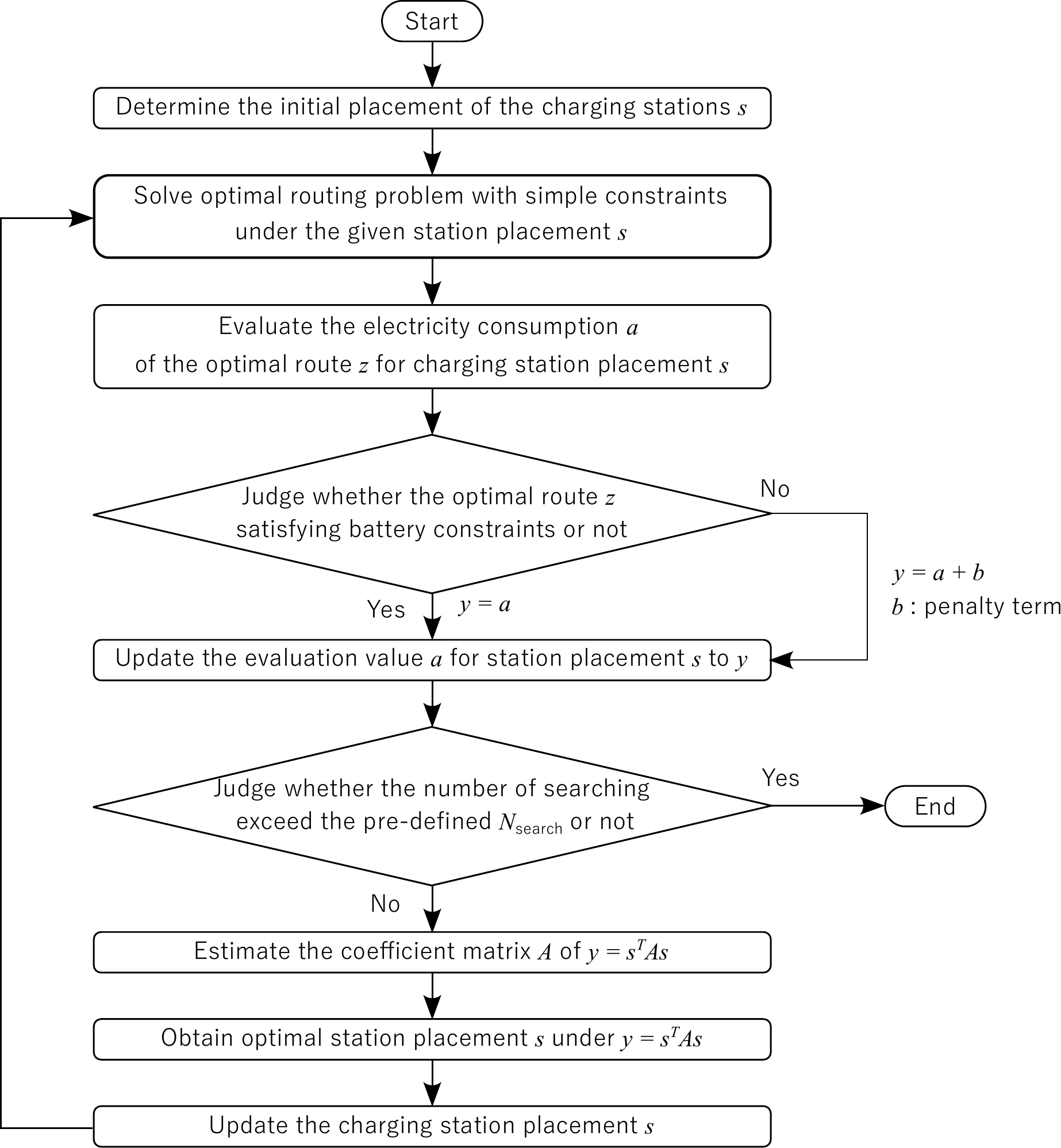}
\caption{
Flowchart of the sequential framework for determining the charging station placement $s$ and the optimal tour $z$.
Given a proposed station configuration $s$, the optimal tour $z$ is obtained by minimizing the energy consumption $a$ without strictly enforcing battery constraints. 
If the resulting tour violates battery constraints, a penalty is added to $a$, and the total becomes the evaluation value $y$. The relationship between the charging station configurations $s$ and their evaluation values 
$y$ is then learned through BOCS.
}
\label{fig:flowchart}
\end{figure}

The overall procedure for jointly optimizing both the placement of charging stations and the vehicle routing under battery constraints is illustrated as a flowchart in Fig.~\ref{fig:flowchart}. 
This procedure consists of proposing charging station configurations using BOCS, followed by searching for an optimal tour that satisfies the battery constraints under the proposed configuration.

\subsection*{Evaluation Data and Parameters}
\begin{figure}[ht]
\centering
\begin{minipage}[t]{0.45\linewidth}
	\centering
	\includegraphics[height=6.5cm]{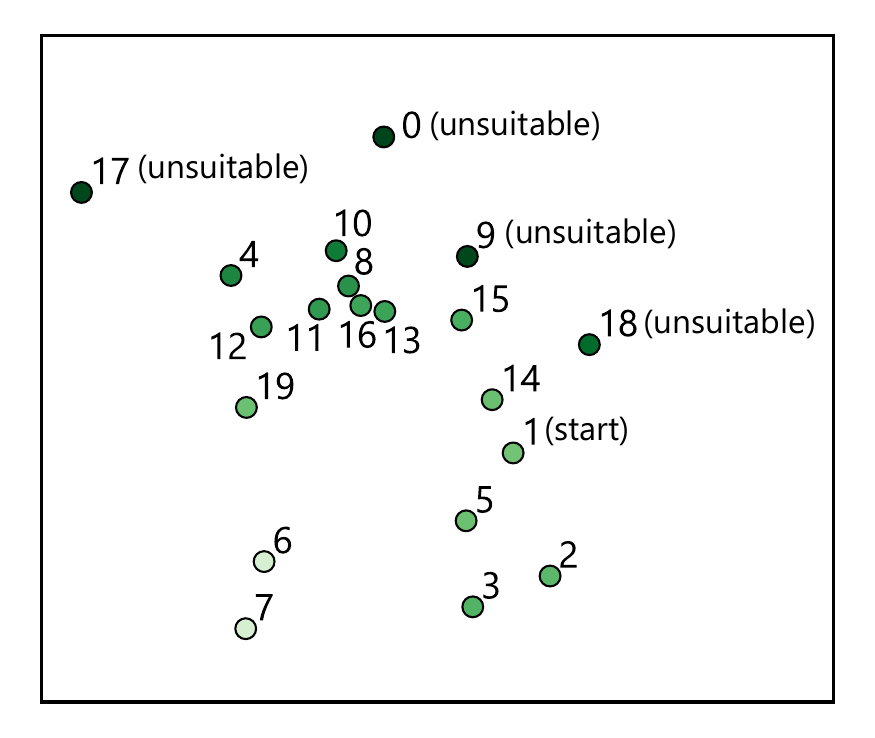}
	\caption{
    Two-dimensional spatial arrangement of the locations used in this study. The numbers next to the markers indicate the location labels. The color represents pseudo-elevation, where higher elevations are assumed to be unsuitable for charging station installation. In this case, locations 0, 9, 17, and 18 are not eligible for station placement.}
	\label{fig:location}
\end{minipage} \hfil
\begin{minipage}[t]{0.45\linewidth}
	\centering
	\includegraphics[height=6.5cm]{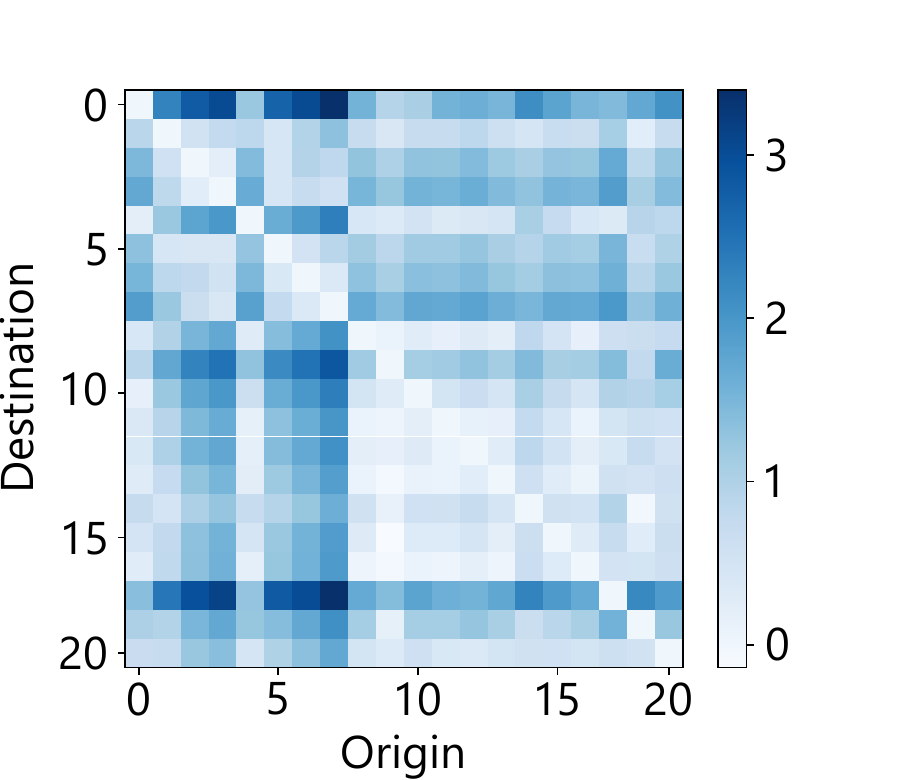}
	\caption{
    Asymmetric travel costs between locations. The values on the $x$- and $y$-axes represent location labels, and the colormap reflects the corresponding travel costs. Some values are negative, representing pseudo-downhill paths where energy regeneration occurs.}
	\label{fig:cost_matrix}
\end{minipage}
\end{figure}

The locations to be visited in the present study are shown in Fig.\ref{fig:location}, and the travel costs between these locations are summarized in Fig.\ref{fig:cost_matrix}.

The parameters used in this study are summarized in Table~\ref{table:parameters}.
Among the 20 candidate locations, 4 are assumed to be unavailable for charging station installation.
The initial $N_\mathrm{Init}$ data samples provided to BOCS consist of randomly generated charging station configurations $s$ and their corresponding evaluation values $y$.
\begin{table}[ht]
\caption{Parameters used in this study}
\centering
\begin{tabular}{c|cc}\hline
Variable & Description & value  \\ \hline
$N$ & Number of locations to visit & 20 \\ 
$M$ & Number of candidate locations for station placement & 16 \\ 
$i_0$ & Starting location (i.e., $x_{i j 1} = 0$ for all $j$ when $i \neq i_0$) & 1 \\
$Q_\mathrm{max}$ & Maximum battery capacity & 6 \\ 
$Q_\mathrm{charge}$ & Battery charge per station visit & 3 \\
$Q_\mathrm{standard}$ & Reference battery level & 3 \\
$N_\mathrm{search}$ & Number of optimization iterations &300 \\
$N_\mathrm{Init}$ & Number of initial data samples & 10 \\ 
$y_\mathrm{penalty}$ & Penalty value for constraint violations & 10 \\ \hline
\end{tabular} \label{table:parameters}
\end{table}

In the routing-QUBO formulation, the penalty coefficients for the one-hot and continuity constraints were set to 
$\lambda_1 = \lambda_2 = \lambda_3 = 400$. 
Although the soft battery-guiding term with coefficient $\lambda_4$ was introduced in Eq.~\eqref{eq:total_guide}, 
it was not activated in the present numerical experiments ($\lambda_4 = 0$). 
Battery feasibility was instead handled exclusively through the post-evaluation penalty within the BOCS framework. 
This setting allows us to isolate the effect of the black-box learning mechanism without introducing additional bias into the routing-QUBO.

To obtain the optimal $s$ that minimizes the learned QUBO function $y = s^T A s$ in BOCS, we employed QA using the Advantage\_system6.3 quantum annealer provided by D-Wave Systems. 
For comparison, we also used Simulated Annealing (SA) included in the Python library ``dwave-ocean-sdk'' provided by D-Wave Systems~\cite{DWS}.
For solving the 20-city routing problem under simplified battery constraints given the charging station configuration proposed by BOCS, we used Fixstars Amplify, a GPU-based annealer.

For the comparison of acquisition optimizers (QA and SA), experiments were repeated using 10 independent random seeds. The convergence behavior was assessed by reporting the mean best-so-far cost together with 95\% confidence intervals computed using the t-distribution. For feasibility rates, Wilson confidence intervals for binomial proportions were employed.

The travel-cost matrix used in this study was derived from a real-world dataset provided by an industrial partner and normalized for confidentiality.
The instances with smaller problem sizes ($N = 10, 13, 15, 18$) were constructed by selecting spatially clustered subsets of nodes from the full dataset.
Therefore, these instances preserve the structural characteristics of the original data and are used primarily for validation and scalability analysis rather than for standardized benchmarking.

\section*{Results}
\subsection*{Small-scale validation by exhaustive enumeration}
To verify that the proposed framework correctly identifies feasible charging-station configurations, we conducted a validation experiment on a smaller instance where the entire configuration space can be evaluated exactly.
Since exhaustive evaluation becomes computationally infeasible for larger instances, this validation experiment is conducted on a reduced problem size in order to verify that the proposed framework correctly identifies globally optimal feasible configurations.
In this experiment, the number of locations to visit was reduced to $N = 10$. 
Charging stations were allowed to be installed at all locations, and the battery parameters were set to $Q_\mathrm{max} = 2.5$ and $Q_\mathrm{charge} = 0.5$. 
The BOCS search was initialized with a single randomly generated configuration. 
Under these conditions, the total number of possible station-placement configurations is $2^{10} = 1024$.
We evaluated all configurations using the same routing solver and evaluator described in Procedure 6.

Among the 1024 configurations, 52 were found to satisfy the battery constraints.
Using this exhaustive dataset, we obtained the globally optimal feasible objective
value and used it as a reference for evaluating the behavior of the proposed
BOCS framework.
\begin{figure}[h]
\centering
\begin{minipage}[t]{0.45\linewidth}
\centering
\includegraphics[width=8.5cm]{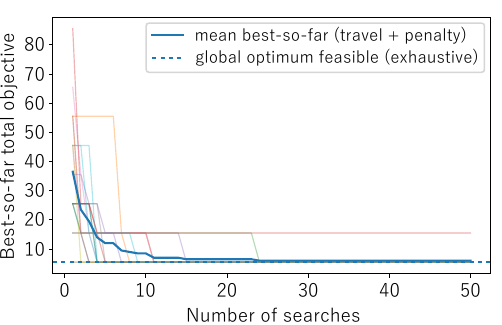}
\caption{
Convergence of the BOCS search compared with the global optimum obtained by exhaustive enumeration for the $N = 10$ validation instance. 
The curve shows the best-so-far objective value (travel cost plus penalty) during the search. The dashed horizontal line indicates the globally optimal feasible value identified from the exhaustive evaluation of all $2^{10}$	charging-station configurations. 
The results demonstrate that the proposed framework rapidly approaches the optimal feasible solution within a small number of iterations.
}
\label{fig:alse_best}
\end{minipage} \hfil
\begin{minipage}[t]{0.45\linewidth}
\centering
\includegraphics[width=8.5cm]{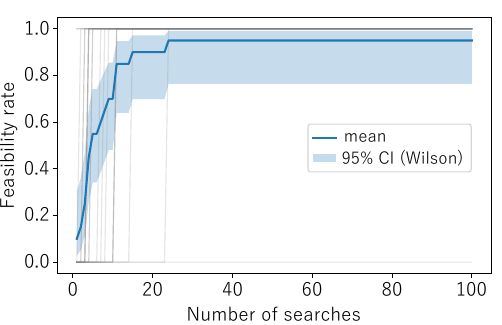}
\caption{
Probability of discovering at least one battery-feasible solution by iteration 
$t$ across $20$ independent runs of the BOCS search for the $N = 10$ validation instance. 
The solid line indicates the empirical probability, and the shaded region represents the Wilson 95\% confidence interval. 
The results show that feasible configurations are typically identified within the early search iterations.
}
\label{fig:alse_feas}
\end{minipage}
\end{figure}
Figure \ref{fig:alse_best} shows the evolution of the best-so-far objective value (travel cost plus penalty) during the BOCS search together with the globally optimal feasible value
obtained by exhaustive enumeration. The results show that the search rapidly
approaches the optimal value. In most runs, a feasible configuration is discovered
within approximately ten search iterations.
Figure \ref{fig:alse_feas} shows the probability that at least one feasible solution has been found by iteration $t$ across independent runs. The shaded region indicates the Wilson
95\% confidence interval. Although one run did not discover a feasible configuration
within the search budget, the majority of runs identify feasible solutions at early
iterations.
It is also noteworthy that even the infeasible candidates identified during the search exhibit objective values very close to the optimal feasible value. 
This indicates that the surrogate model learned by BOCS guides the search toward regions near the feasible boundary.
These results confirm that the proposed black-box learning framework is capable of identifying feasible and near-optimal station configurations when the global optimum is known.

\subsection*{Search performance}
\begin{figure}
\centering
\begin{minipage}[t]{0.48\linewidth}
	\centering
	\includegraphics[width=8cm]{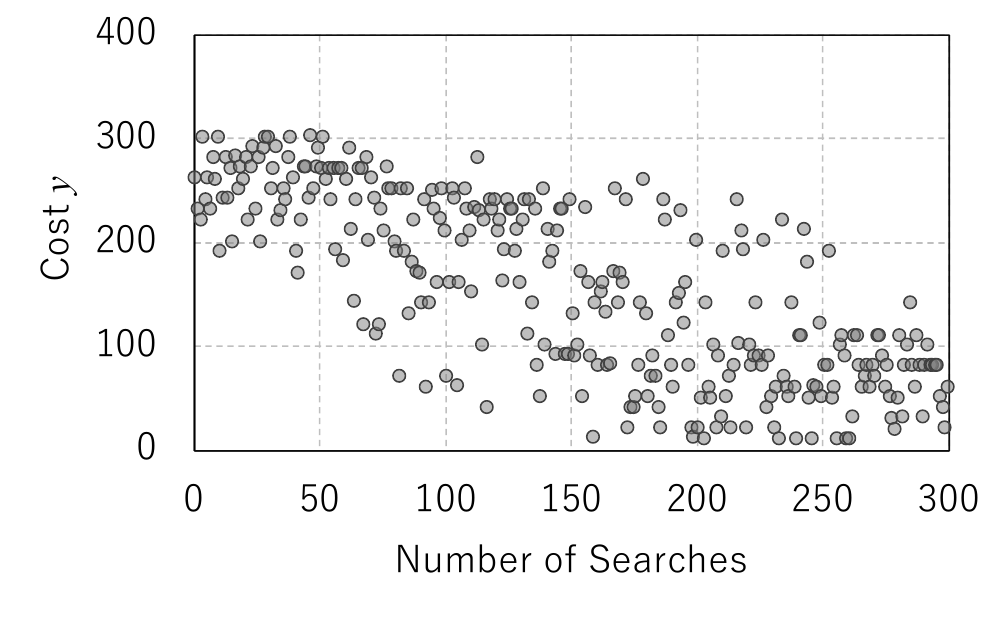}
	\label{fig:all_searcha}
\end{minipage}
\begin{minipage}[t]{0.48\linewidth}
    \centering
    \includegraphics[width=8cm]{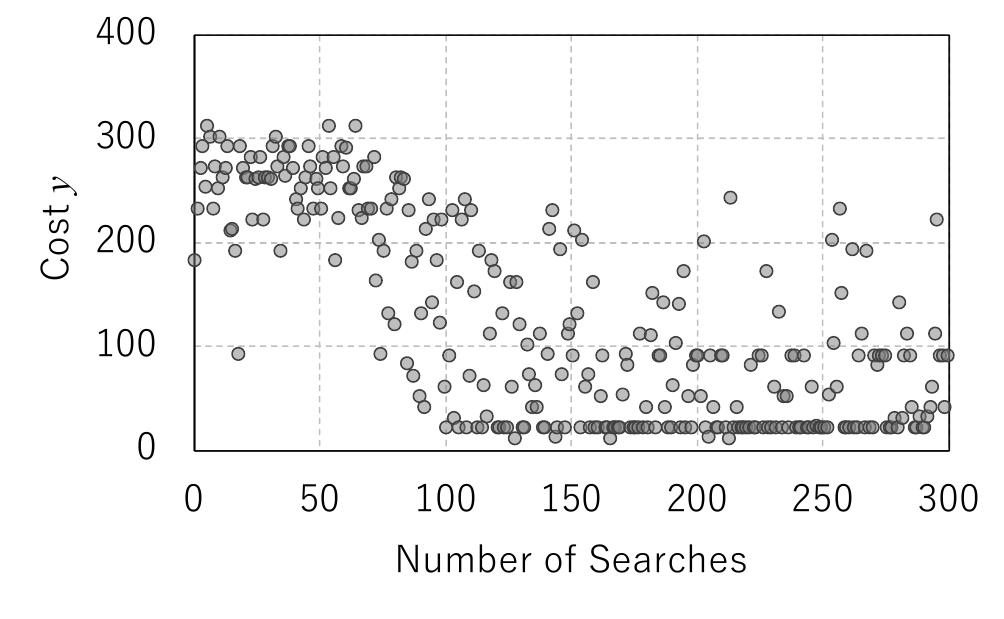}
    \label{fig:all_searchb}
\end{minipage}
\caption{Representative search trajectories of BOCS-QA under different random seeds. The vertical axis labeled 'Cost' represents the final evaluation value $y$ from Procedure 6. While the learning dynamics differ across runs, it is consistently observed that the search becomes more efficient in identifying high-quality solutions as the number of iterations increases.}
\label{fig:all_searches}
\end{figure}

Next, we analyze the relationship between the number of search iterations and the corresponding evaluation values $y$. Fig.~\ref{fig:all_searches} presents search histories obtained under different random seeds.
In the left panel, the search initially yields many solutions that violate battery constraints, and such infeasible solutions continue to appear intermittently even after 200 to 250 iterations. In contrast, the right panel shows a more rapid convergence toward feasible regions, with constraint violations significantly reduced by around 100 to 150 iterations. Nevertheless, despite these differences in convergence behavior and exploration dynamics, both cases ultimately achieve frequent identification of feasible, low-cost solutions that satisfy battery constraints.

\subsection*{Comparison Quantum annealing with Simulated annealing}
\begin{figure}
\centering
\includegraphics[width=8.5cm]{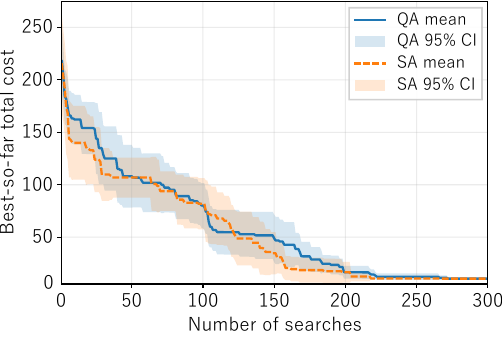}
\caption{
Evolution of the best-so-far tour cost over search iterations.
The curves show the mean best-so-far cost averaged over 10 independent runs. The shaded regions indicate 95\% confidence intervals computed using the t-distribution. Although Quantum annealing (QA) attains slightly lower mean costs in later iterations, the confidence intervals overlap substantially, suggesting no statistically significant difference between acquisition optimizers under the present experimental setting.
}
\label{fig:record}
\end{figure}

To quantitatively assess convergence behavior, we analyzed the evolution of the best-so-far tour cost across independent runs.
Fig.~\ref{fig:record} shows the mean best-so-far cost over 10 random seeds as a function of the search iteration. The results obtained using Quantum Annealing (QA) are shown with solid line, and those from Simulated Annealing (SA) are shown with dashed line. The shaded regions represent 95\% confidence intervals computed using the t-distribution.
A detailed comparison of the tour costs at the 300th iteration is provided in Table~\ref{table:results_compare}. 
While minor differences in convergence speed are observed, the confidence intervals largely overlap throughout the search. Therefore, under the present experimental setting, no statistically significant superiority can be claimed for either acquisition optimizer in terms of cost convergence.
\begin{table}[ht]
\caption{Tour Cost Comparison at the 300th Iteration using SA and QA}
\centering
\begin{tabular}{c|cccc}\hline
Method & Worst & Best & Mean & Variance \\ \hline
SA & 11.17 & 9.74 & 10.50 & 0.40 \\ 
QA & 11.38 & 9.53 & 10.43 & 0.50 \\ \hline
\end{tabular} \label{table:results_compare}
\end{table}

\begin{figure}
	\centering
	\includegraphics[width=8.5cm]{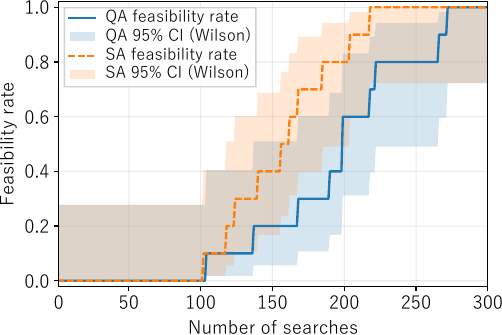}
	\caption{{
			Feasibility rate of the best-so-far solutions as a function of iteration, averaged over 10 independent runs. The rate indicates the proportion of runs whose best-so-far solution satisfies the battery constraints at each iteration. Shaded regions denote 95\% Wilson confidence intervals. This metric evaluates cumulative best-so-far feasibility rather than instantaneous candidate feasibility.}}
	\label{fig:feasi}
\end{figure}

In addition to cost reduction, it is important to examine whether the search process consistently identifies battery-feasible solutions. Therefore, we next analyze the feasibility status of the best-so-far solutions.
For each of the 10 independent runs, we recorded whether the best solution obtained up to iteration $t$ satisfied the battery constraints. 
The feasibility rate shown in Figure \ref{fig:feasi} represents the proportion of runs (out of $10$) whose best-so-far solution is feasible at iteration $t$, resulting in a resolution of $0.1$. Shaded regions indicate $95\%$ Wilson confidence intervals for binomial proportions.

As the search progresses, the feasibility rate increases monotonically, reflecting the learning behavior of the BOCS framework. At the final iteration ($t$ = 300), both QA and SA achieved a feasibility rate of 1.0 (10/10 runs), indicating that the framework consistently identifies battery-feasible solutions. Although SA appears to reach high feasibility slightly earlier in some runs, the confidence intervals largely overlap, suggesting no statistically significant difference between acquisition optimizers in terms of constraint satisfaction.

\subsection*{Best solution analysis}
\begin{figure}
\centering
\includegraphics[width=10cm]{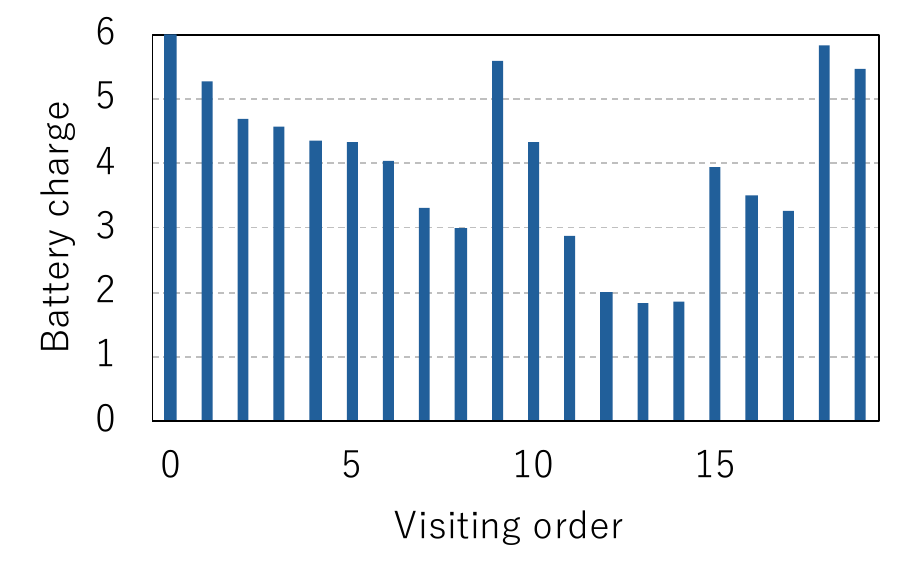}
\caption{Battery level history during the tour that yielded the best tour cost under QA.}
\label{fig:battery_history}
\end{figure}

\begin{figure}
\centering
\includegraphics[width=12cm]{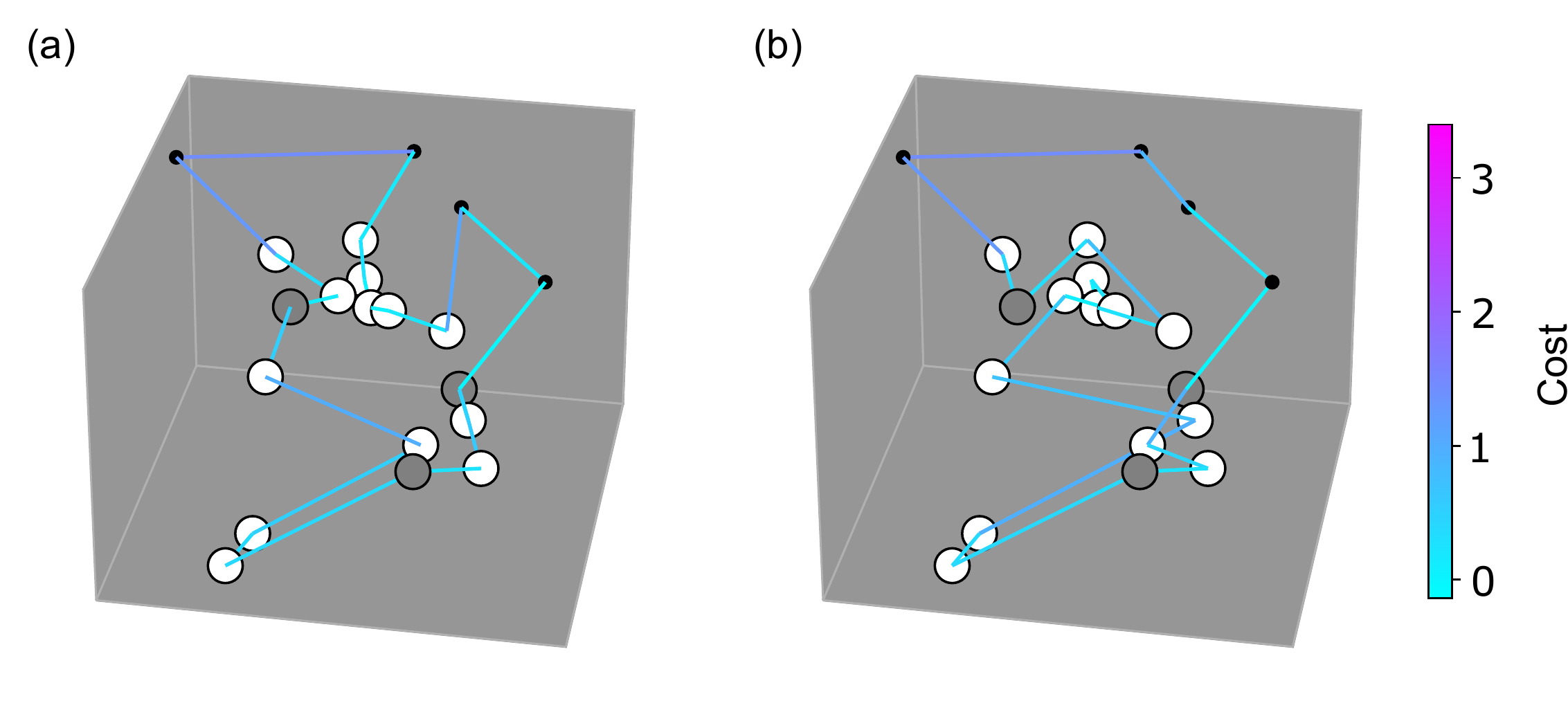}
\caption{Effect of battery constraints on routing.
(a) Route obtained by minimizing travel cost only.
(b) Route obtained while considering battery level constraints.
The black dots indicate locations where charging stations cannot be installed, while the gray circles represent locations where charging stations are installed, and the white circles indicate unselected locations.}
\label{fig:TSP_Route}
\end{figure}

Next, we describe the solution that yielded the best evaluation value, focusing on the state interpreted from the corresponding binary variables.
In the case of QA, the optimal configuration of charging stations was at locations 3, 12, and 14. 
The visiting order of the locations was:
$1 \to 19 \to 11 \to 13 \to 8 \to 16 \to 15 \to 10 \to 12 \to 4 \to 17 \to 0 \to 9 \to 18 \to 14 \to 5 \to 2 \to 3 \to 7 \to 6$.
The battery level history along this route is shown in Fig.~\ref{fig:battery_history}. 
As can be seen, the battery constraint was indeed satisfied throughout the tour.
The tour obtained while considering battery constraints is shown in Fig.~\ref{fig:TSP_Route}. For comparison, the figure also includes the route derived by minimizing only the travel cost, without accounting for battery constraints.

\subsection*{Computational resource usage}
To evaluate the computational characteristics of the proposed framework, we analyze the wall-clock runtime for a full optimization run after the 300 search iterations. 
Since the runtime per iteration fluctuates due to stochastic solver behavior, caching effects, and hardware scheduling, we report the aggregated runtime over the entire search process, which represents the effective computational unit of the method.

\begin{figure}
\centering
\includegraphics[width=8.5cm]{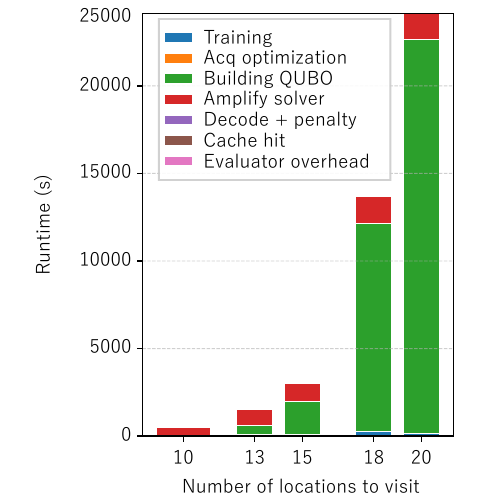}
\caption{
Wall-clock runtime breakdown for a full optimization run consisting of $N_{\mathrm{search}}=300$ iterations.
The total runtime is decomposed into BOCS model training (Training), acquisition optimization (Acq optimization),
routing-QUBO construction (Building QUBO), routing-QUBO solving via Fixstars Amplify (Amplify solver),
solution decoding and battery-penalty evaluation (Decode + penalty), cache reuse (Cache hit),
and additional evaluator overhead.
For $N=20$, the number of candidate charging-station locations was fixed to $M=16$ to match the QA--SA comparison setting.
The results indicate that routing-QUBO construction and solving dominate the computational cost,
while the surrogate learning component remains comparatively small across problem sizes.
}
\label{fig:runtime}
\end{figure}

Figure \ref{fig:runtime} presents the runtime decomposition for problem sizes $N = 10, 13, 15, 18$, and $20$.
The total runtime is divided into the following components: (i) BOCS model training (Training), (ii) acquisition optimization for determining the next charging station configuration (Acq optimization), (iii) construction of the routing-QUBO (Building QUBO), (iv) routing-QUBO solving using the Amplify GPU-based annealer (Amplify solver), (v) decoding of the binary solution and evaluation of battery constraints including penalty calculation (Decode + penalty), (vi) cache reuse (Cache hit), and (vii) additional evaluator overhead.
As shown in Fig. \ref{fig:runtime}, the dominant computational cost arises from the construction and solving of the routing-QUBO, particularly for larger values of 
$N$. In contrast, the time required for BOCS model training and acquisition optimization constitutes only a small fraction of the total runtime across all tested problem sizes. 
This indicates that the black-box learning component does not become a computational bottleneck in the current framework. 
The observed growth in runtime with increasing $N$ is primarily attributable to the increase in routing-QUBO size, which leads to higher construction cost and longer annealing time.
For the case $N = 20$, only 16 locations were considered as candidates for charging station installation, whereas in the other problem sizes all locations were eligible. 
However, because the total runtime is dominated by the routing-QUBO construction and solving phases rather than by the dimensionality of the BOCS search space, this difference in candidate set size has a negligible impact on the overall computational profile.
These results clarify the computational structure of the proposed method and identify routing-QUBO generation and solution as the primary targets for future scalability improvements.

All experiments were conducted on a workstation running Ubuntu 24.04 LTS, equipped with an Intel Xeon Platinum 8488C CPU (16 logical cores) and 64 GB RAM. 
The routing-QUBO was solved using the Fixstars Amplify solver (version 1.3.1) via cloud access to GPU-based computational resources. 
Acquisition optimization was performed using the D-Wave Ocean SDK (version 8.3.0) with access to Advantage\_system6.3. Python 3.12.9 was used for all experiments.

\subsection*{QUBO size analysis}
To further clarify the computational characteristics of the proposed framework, we analyzed the size of both the routing-QUBO and the surrogate QUBO learned by BOCS.

The routing formulation introduces $N^3$ binary variables $x_{ijt}$. 
After expanding the quadratic penalties used in the present experiments, 
the number of nonzero coefficients grows approximately as $O(N^5)$.
If the soft battery-guidance term is activated, the QUBO may become denser, 
and the number of quadratic coefficients can scale up to $O(N^6)$.
For example, when $N=20$, the routing-QUBO consists of 8{,}000 binary variables and 1{,}824{,}000 nonzero coefficients (8{,}000 linear and 1{,}816{,}000 quadratic terms) as shown in Table~\ref{tab:qubo_size}. 
This rapid growth explains why QUBO construction and solving dominate the runtime for larger problem sizes.

The surrogate model learned by BOCS is a quadratic function of the station-placement variables $s_i$,
\begin{equation}
y = \alpha_0 + \sum_i \alpha_i s_i + \sum_{i<j} \alpha_{ij} s_i s_j.
\end{equation}
For $M$ candidate station locations, the surrogate model contains $M$ binary variables and 
$1 + M + \binom{M}{2}$ coefficients, i.e., $O(M^2)$ complexity.
In the experiments for $N=20$, we fixed the number of candidate station locations to $M=16$ to maintain consistency with the QA–SA comparison setting. 
Under this configuration, the surrogate-QUBO contains 16 binary variables and 137 coefficients 
(16 linear and 120 quadratic terms, plus one intercept). 
Therefore, the surrogate model is significantly smaller than the routing-QUBO and does not dominate the computational cost.

As shown in Table~\ref{tab:qubo_size}, the surrogate model remains substantially smaller than the routing-QUBO across all problem sizes. 
These results confirm that the primary scalability bottleneck of the proposed framework arises from the routing-QUBO rather than from the black-box learning component.
\begin{table}[t]
\centering
\caption{
Size of routing-QUBO and BOCS surrogate-QUBO for different problem sizes. 
For $N=20$, the number of candidate station locations was fixed to $M=16$ to match the QA–SA comparison setting.
}
\label{tab:qubo_size}
\begin{tabular}{c|c|c|c|c}
		\hline
		$N$ & Routing variables & Routing nonzeros & $M$ & Surrogate coefficients \\
		\hline
		10 & 1,000 & 63,000 & 10 & 56 \\
		13 & 2,197 & 224,094 & 13 & 92 \\
		15 & 3,375 & 448,875 & 15 & 121 \\
		18 & 5,832 & 1,090,584 & 18 & 172 \\
		20 & 8,000 & 1,824,000 & 16 & 137 \\
		\hline
	\end{tabular}
\end{table}

\section*{Discussion}
This section first discusses the obtained results, then examines the applicability of the proposed framework to constrained optimization problems that are difficult to explicitly formulate in QUBO form, and finally provides concluding remarks.

We now discuss the results obtained in this study.
The relationship between the number of search iterations and the evaluation values suggests that the proposed framework is effective for solving constrained optimization problems.
As shown in Fig.\ref{fig:all_searches}, the progression of learning varies depending on the random seed. Nevertheless, the framework was able to identify high-quality, constraint-satisfying solutions within 300 iterations (Fig. \ref{fig:record}).
Considering that the number of possible charging station configurations is 
$2^{16}$, 
the ability to obtain feasible high-quality solutions within 300 iterations indicates stable convergence behavior in the tested setting.

The resulting arrangement of charging stations and the corresponding tour routes were also deemed reasonable.
The charging stations were located at mutually distant positions, naturally eliminating plans that would require frequent recharging. 
Moreover, the derived routes differed from those that simply minimize travel costs, indicating that the placement of charging stations was appropriately taken into account.
However, according to the battery history shown in Fig.~\ref{fig:battery_history}, it appears that the final charging station (Station 3) could be omitted without violating any constraints. 
Therefore, if the installation of charging stations is associated with a cost, it would be appropriate to incorporate a penalty term related to the number of stations into the objective function $y=s^\top A s$ during the search for the next candidate solution after training with BOCS.

With regard to the performance of QUBO solvers, there was no significant difference between Simulated Annealing (SA) and Quantum Annealing (QA) in the context of the example considered in this study.
Although QA achieved marginally lower mean best-so-far costs, the overlapping confidence intervals indicate that no statistically significant difference can be established under the present experimental setting. This absence of statistical significance may partly stem from the relatively small problem size and the simplicity of the approximated QUBO formulation, which both acquisition optimizers can solve effectively. Therefore, the present results should not be interpreted as evidence against potential advantages of QA in larger or more complex instances.

{While QUBO and annealing-based solvers are used in this study, the proposed framework is not restricted to quantum annealing itself. 
Instead, it relies on a QUBO formulation combined with a quadratic surrogate model learned via BOCS. 
Within this structure, different QUBO solvers (e.g., QA, SA, or GPU-based annealers) can be used as interchangeable optimization backends. 
Therefore, the main contribution of this work lies in the surrogate-based handling of implicit constraints within a QUBO-compatible formulation, rather than in demonstrating a quantum advantage.}

We next examine the applicability of the proposed framework to constrained optimization problems that are difficult to formulate explicitly in QUBO form.
Although it is not possible to formally explain why the proposed framework effectively functions as a method for optimal route planning under battery constraints and for the optimal placement of charging stations, 
two possible contributing factors can be considered.

The first factor is that the expressive power of the QUBO formulation may be sufficiently high in the vicinity of the optimal solutions. The proposed framework focuses the search around the neighborhood of the optimal solutions, and the learning progresses in QUBO form accordingly. Even if the original objective function includes constraints and is somewhat complex, it is plausible that - similar to a Taylor expansion of a continuous function - such a function can be adequately approximated by a QUBO formulation within the vicinity of the optimum. This notion is supported by the surprisingly broad applicability of black-box optimization using QUBO formulations across various domains.

The second potential contributing factor is the role of the soft battery-guidance term defined by $Q_{\mathrm{standard}}$ in the routing-QUBO formulation. 
In the numerical experiments reported in this study, this term was not activated ($\lambda_4 = 0$). 
Instead, battery feasibility was enforced solely through the post-evaluation penalty within the BOCS framework. 
This setting was intentionally adopted to isolate the effect of the black-box learning mechanism and to avoid introducing additional bias into the routing-QUBO during the search process.
Nevertheless, the soft battery-guidance term provides an intuitive mechanism for incorporating domain knowledge about battery dynamics into the routing formulation. 
Without such a guiding term, the routing optimization may focus exclusively on minimizing travel cost, potentially leading the search toward routes that systematically violate battery constraints. 
By encouraging intermediate battery levels around a reference value $Q_{\mathrm{standard}}$, the formulation can bias the search toward routes that are more likely to remain within feasible battery ranges.
Moreover, this guidance may indirectly influence the placement of charging stations by discouraging solutions that rely on excessive or poorly distributed charging locations. 
Even when charging stations are redundantly placed, the routing-QUBO still attempts to generate routes that maintain reasonable battery levels throughout the tour. 
Consequently, the evaluation process may return objective values that reflect natural route quality rather than artificial penalties caused by large constraint violations, thereby promoting more informative learning near the feasible boundary of the solution space.
Although the present experiments focused on the penalty-based feasibility evaluation, future studies may investigate the impact of activating the soft battery-guidance term ($\lambda_4 > 0$) on convergence speed, exploration behavior, and solution quality.

We further clarify the methodological relationship between the proposed framework and related surrogate-based approaches for combinatorial problems such as COMBO~\cite{Oh2019} and FMQA~\cite{Kitai2020}. 
Although these methods share the paradigm of sequential surrogate learning and combinatorial acquisition optimization, their surrogate structures differ.
COMBO employs Gaussian process models defined over combinatorial domains. While this enables flexible modeling and uncertainty quantification, the surrogate model itself is not restricted to quadratic forms and therefore is not directly represented as a QUBO.
Therefore, if QUBO solvers are used as the primary optimization engine, additional transformation steps would be required.
In contrast, FMQA utilizes factorization machines as surrogate models and is designed to work naturally with QUBO solvers during acquisition optimization. 
From a structural viewpoint, the overall sequential framework adopted in this study could also be implemented using FMQA without major modification.
The principal difference between the present approach and FMQA lies in the treatment of uncertainty and exploration. BOCS provides a Bayesian inference scheme tailored to quadratic surrogate models, allowing posterior sampling of model parameters and uncertainty-aware exploration within the QUBO-structured search space. 
In contrast, FMQA typically relies on point estimation of surrogate parameters. The Bayesian mechanism adopted in this study may enhance exploration efficiency by balancing exploitation and uncertainty-driven sampling. On the other hand, the additional modeling assumptions and sampling procedures may introduce computational overhead or limit expressive power compared with more flexible surrogate models. Whether Bayesian exploration offers a systematic advantage in electric vehicle routing with charging constraints remains an empirical question and represents an important direction for future work.
{
From the perspective of model interpretability, while the surrogate model learned by BOCS plays a central role in the proposed framework, its internal structure and the manner in which constraint-related information is implicitly captured are not explicitly analyzed in the present study. In particular, the extent to which constraint-related penalties are disentangled from the underlying routing cost in the learned surrogate model remains unclear. A more detailed interpretation of the surrogate model would provide further insight into the learning mechanism and represents an interesting direction for future work.}

The approach proposed in this study --- utilizing a QUBO solver for constrained combinatorial optimization --- also has certain limitations.
The method was specifically applied to a constraint that requires the battery level to remain within a fixed range. 
While the framework proved effective for this particular setting, extending it to other types of constraints may require certain modifications or adjustments.
The general idea of introducing penalty terms for constraint violations is broadly applicable and, in principle, can be extended to various constraint conditions.
However, in this work, the vehicle routing formulation included an additional term that encourages the battery level to remain near a reference value $Q_\mathrm{standard}$, leveraging domain-specific knowledge unique to electric vehicle operation.
Even for other range-type constraints --- such as ensuring deliveries are completed within specific time windows --- the same formulation may not be directly transferable.
This aspect represents a limitation of the current study and highlights a promising direction for future research.

From the perspective of future work, various extensions of the proposed framework can be considered.
For example, relationships between charging stations—such as tendencies to be co-located or to be mutually exclusive—can be incorporated into the optimization process by adding corresponding penalty terms to the objective function learned through BOCS. This would allow such inter-station relationships to be taken into account when determining optimal charging station placements.

{A further limitation of the present study lies in the formal verification of the route-encoding QUBO formulation. 
Although the validity of the overall framework is supported by the exhaustive enumeration experiments, 
these results do not fully isolate the correctness of the QUBO encoding itself. 
In particular, a formal proof of equivalence to the standard vehicle routing formulation or validation through exact methods such as MILP has not been provided in this work. 
Addressing this point would require a more detailed theoretical analysis or systematic cross-verification, 
which remains an important direction for future research.}

In conclusion, this study addressed the simultaneous optimization of electric vehicle routing with battery constraints and the placement of charging stations.
Given a configuration of charging stations, we formulated the vehicle routing problem—considering battery capacity in a simplified manner—as a QUBO model, which is compatible with quantum annealing, and used it to determine the corresponding tour.
To optimize the placement of charging stations, we employed BOCS, a sequential optimization method. The tour cost used for learning was computed based on the routing results obtained from the QUBO formulation.
As a result, the BOCS framework effectively learned the impact of battery constraints through its sequential search process, enabling efficient exploration and yielding solutions that satisfied all constraints.
Validation experiments based on exhaustive enumeration and scalability analyses clarify both the correctness and computational characteristics of the proposed framework.

In summary, the key contributions of this study can be organized into the following three points.
\begin{enumerate}
    \item We proposed a novel black-box optimization framework that enables joint optimization of charging station placement and vehicle routing under battery constraints without requiring an explicit formulation of such constraints in QUBO form.
    \item By combining a soft constraint formulation with BOCS, we demonstrated that high-quality, feasible solutions can be obtained efficiently even in complex, constrained combinatorial problems where standard QUBO formulations face scalability issues.
    \item {Our method provides a complementary strategy for leveraging QUBO solvers in constrained combinatorial optimization, particularly in settings where explicit constraint modeling leads to large and dense formulations.}
\end{enumerate}
Rather than replacing classical operations research (OR) techniques such as mixed-integer linear programming (MILP) or advanced metaheuristics, the proposed framework demonstrates how black-box learning combined with QUBO solvers can complement existing approaches, particularly in settings where explicit constraint modeling becomes complex or leads to large and dense formulations under hardware limitations.
These contributions suggest a promising direction for extending QUBO-based approaches to other real-world optimization problems that involve implicit or hard-to-model constraints.

\section*{Data availability}
The data used and analyzed during the current study are available from the corresponding author on reasonable request.

\section*{Funding}
This research received no external funding.

\section*{Author information}
\subsection*{Contributions}
A.O. proposed basic concept, K.O. and H.Y. proposed formulations, 
A.O. and K.O. conducted the numerical experiments. 
All authors analysed the results and reviewed the manuscript. 

\subsection*{Corresponding author}
Correspondence to Akihisa Okada.

\section*{Competing interests}
The authors declare that this work is related to patent applications and a granted patent (Japanese Patent Application No. 2023-130687; Japanese Patent Publication No. 2025-025669; Japanese Patent No. 7800517). The authors declare no other competing interests. The authors confirm that these interests did not influence the design, analysis, or reporting of the study.

\bibliography{LearningChargeConstraints_0302}

\end{document}